\documentclass[%
 reprint,
superscriptaddress,
nobibnotes,
 amsmath,amssymb,
 aps,
prb,
citeautoscript,
floatfix,
showkeys
]{revtex4-2}
\usepackage[T1]{fontenc} % Use modern font encodings
\usepackage[]{graphicx}
\usepackage{bm}
\usepackage[normalem]{ulem}
\usepackage{amssymb}
\usepackage{amsmath}
\setcitestyle{super}

\newcommand{\dqmp}{Department of Quantum Matter Physics, University of Geneva, 24 Quai Ernest Ansermet, CH-1211 Geneva, Switzerland}
\newcommand{\gap}{Department of Applied Physics, University of Geneva, 24 Quai Ernest Ansermet, CH-1211 Geneva, Switzerland}

\begin{document}

	%	\preprint{v6}

	\title{Full control of solid-state electrolytes for electrostatic gating}%

	\author{Chuanwu Cao}             %fan.wu@unige.ch
	\author{Margherita Melegari}
 	\author{Marc Philippi}
  	\author{Daniil Domaretskiy}
   	\author{Nicolas~Ubrig}% gibertini@unimore.it
    \author{Ignacio Gutiérrez-Lezama}      % ignacio.gutierrez@unige.ch
    \author{Alberto F. Morpurgo}
    \email{alberto.morpurgo@unige.ch}
    \affiliation{\dqmp}
    \affiliation{\gap}

	\date{\today}% It is always \today, today,

\begin{abstract} %200 words max
Ionic gating is a powerful technique to realize field-effect transistors (FETs) enabling experiments not possible otherwise. So far, ionic gating has relied on the use of top-electrolyte gates, which pose experimental constraints and make device fabrication complex. Promising results obtained recently in FETs based on solid-state electrolytes remain plagued by spurious phenomena of unknown origin, preventing proper transistor operation, and causing  limited control and reproducibility. Here we explore a  class of solid-state electrolytes for gating (Lithium-ion conducting glass-ceramics, LICGCs), identify the processes responsible for the spurious phenomena and irreproducible behavior,and demonstrate properly functioning transistors exhibiting high density ambipolar operation with gate capacitance  of $\approx 20-50\: \mu\mathrm{F/cm^2}$ (depending on the polarity of the accumulated charges).  Using two-dimensional semiconducting transition-metal dichalcogenides we demonstrate the ability to implement ionic-gate spectroscopy to determine the semiconducting bandgap, and to accumulate electron densities  above 10$^{14}$ cm$^{-2}$, resulting in gate-induced superconductivity in MoS$_2$ multilayers. As LICGCs are implemented in a back-gate configuration, they leave the surface of the material exposed, enabling the use of surface-sensitive techniques (such as scanning tunneling microscopy and photoemission spectroscopy) impossible so far in ionic-liquid gated devices. They also allow double ionic gated devices providing independent control of charge density and electric field.
\end{abstract}

	\maketitle

\section{Introduction}

%\subsection{First Subsection}
%\subsubsection{First Sub Subsection}
%\threesubsection{First lowest-level subsection}

%\textcolor{red}{added or modified text}
%\textcolor{blue}{\sout{deleted text}}
%\textcolor{blue}{\sout{}}

Ionic gates exploit the motion of ions in an electrolyte to transfer the potential applied to a metallic gate electrode to the surface of a material\cite{brattain_experiments_1955,EDL_Review_2013,Frisbie_review_2013,IL_Review_2017,ILS_review_2021,torricelli2021electrolytebio}. Their efficiency originates from the very large geometrical capacitance $C_\mathrm{G}$ (10-50 $\mathrm{\mu F/cm^2}$\cite{lee2007ion,Geo-IL_2016,IL_Review_2017,CrossQuantum_2019}) of the electric double layer (EDL), less than one nanometer thick, which forms at the interface between the electrolyte and the gated material\cite{EDL_Review_2013,Frisbie_review_2013,IL_Review_2017}. In transistors, the large capacitance leads to unrivaled performance in low-voltage operation\cite{shimotani2006electrolyte,panzer2007polymer,cho2008printable} and sub-threshold swing\cite{cho2008printable,LET_2014_M,ILS_review_2021}. It also enables the observation of physical phenomena, such as gate-induced superconductivity (at carrier densities $\sim 10^{14}\:\mathrm{cm^2}$)\cite{ye2010liquid,ueno2011discovery,MoS2_SC_2012,WS2_SC_2015,ZrNCl_SC_2015,TMDSC_2015,MoS2_SC_2016,IL_Review_2017,costanzo2018tunnelling,lu2018full,piatti2018multi,kouno2018superconductivity,zeng2018gate,piatti2019ambipolar}, the quenching of the band gap of two-dimensional (2D) semiconductors \cite{DoubleGating_2022,DoubleGating_2022_Bolotin}, or the possibility to perform precise spectroscopic measurements of semiconducting band gaps (ionic-gate spectroscopy)\cite{Spectroscopy_2012,MoTe2_2014,FuS_PEO_2015,BP_2015,ReS2_2016,Kis_2018,SWNT_2018,MoS2_2018,artificial_2018,CrossQuantum_2019,ILS_review_2021}.  \\

These results have been enabled by liquid electrolytes, either ionic liquids and ionic solutions (ions dissolved in liquid electrolytes), or ion gels, i.e., ionic liquids or ions dispersed in a polymer matrix (such as polyethylene oxide), ideal for gating applications\cite{IL_Review_2017} because of their high electrochemical stability\cite{EDL_Review_2013,Frisbie_review_2013,IL_Review_2017} and relatively large ionic conductivity\cite{Frisbie_review_2013,IL_Review_2017,DEME-TFSI_Condivity}. As commonly employed in transistors devices, however, these electrolytes impose important constraints. They bury the surface of the gated material, preventing the use of all surface sensitive experimental probes for the investigation of electronic phenomena occurring in the transistor channel. They also limit the realization of all but the simplest device structures, since no additional fabrication steps can be done after drop casting the liquid. \\

Developments in the field of 2D materials are making it clearly apparent that the ability to perform ionic liquid gating in a back-gate configuration has an enormous potential for different experiments that are currently impossible. For instance, a back-gate geometry is essential to perform angle-resolved photoemission studies (ARPES) on exfoliated layers of 2D materials with a variable charge density. ARPES experiments of this type have been performed recently for the first time, but gating was only possible by employing conventional solid-state dielectrics, limiting the maximum charge density to less than $10^{13}$ cm$^{-2}$\cite{nguyen2019visualizing,joucken2019visualizing}. Electrolyte back-gated devices would drastically expand the range of carrier density, and allow phenomena such as gate-induced superconductivity to be studied by ARPES (or, similarly, by scanning tunneling spectroscopy). Ionic double gated devices --which enable the applications of electric fields in excess of 3 V/nm perpendicular to 2D semiconductors to quench their gap-- provide another pertinent and timely example\cite{DoubleGating_2022,DoubleGating_2022_Bolotin}, since they require combining an top and back ionic-gate electrodes. These and other types of experiments cannot be performed using exclusively top ionic-gate electrodes, and to make them possible it is necessary to find solid-state electrolytes (SSE)\cite{knauth2009inorganic,manthiram2017lithium,famprikis2019fundamentals,zheng2018review} that can be employed reliably in a back-gate configuration, with performance comparable to that of top ionic gates.\\

A handful of  experiments on transistors indicate that conductive glass-ceramics containing alkali ions, Li$^+$ or Na$^+$, are promising candidates for electrostatic gating\cite{Glass_SC_2015,Gra_LICGC_2016,Sodium_2017,Marc_LICGC,WS2_LICGC_2020,MoS2_LICGC_2020,MoS2_LICGC_2021} (we are not referring here to experiments where the glass-ceramics are used as a source for ion intercalation\cite{Inter_SnSe_2019,Inter_CGT_2021,Inter_TiSe2_2021}). Back-gated devices employing glass-ceramic substrates, in combination with graphene or transition-metal dichalcogenide (TMD) semiconducting monolayers, were shown to exhibit ambipolar transport\cite{Gra_LICGC_2016,Sodium_2017,WS2_LICGC_2020} and to operate as light-emitting transistors\cite{Marc_LICGC}. There are, however, troublesome inconsistencies between results reported by different groups\cite{Glass_SC_2015,Marc_LICGC,MoS2_LICGC_2020,MoS2_LICGC_2021,Sodium_2017,WS2_LICGC_2020}. A crucial one is the value of the geometrical gate capacitance of conducting glass-ceramic gates, ranging from 1-2 $\mathrm{\mu F/cm^2}$ to more than 100 $\mathrm{\mu F/cm^2}$ in different reports\cite{Sodium_2017,MoS2_LICGC_2020,MoS2_LICGC_2021} (i.e., from 50 times smaller to 2-3 times larger than the geometrical capacitance of commonly used ionic liquids). The situation is especially problematic for electron accumulation, with maximum reported densities ranging from $\sim 5\times 10^{12}\:\mathrm{cm^{-2}}$\cite{MoS2_LICGC_2020,MoS2_LICGC_2021} to 5-6$\:\times\: 10^{14}\:\mathrm{cm^{-2}}$\cite{Sodium_2017} in different systems, a spread that cannot be accounted for by the difference in quantum capacitance of the gated materials, and with multiple experiments showing behavior different from that expected from a field-effect transistor (FET). This includes the absence of charge accumulation --or the saturation of the conductivity at very low values-- when applying positive gate voltages $V_\mathrm{G}$, an extremely slow reaction of the ceramic gate to the applied gate voltage (and hence of the accumulated charge), and the observation of high contact resistance, which is atypical in ionic-gated FETs \cite{Glass_SC_2015,Marc_LICGC,WS2_LICGC_2020}. As a result, in none of the devices reported thus far the key results that make ionic-liquid gated transistor interesting could be reproduced, neither gate-induced superconductivity  nor ionic-gate spectroscopy. Drastic advances in understanding and control are clearly needed to reliably operate  ionic gates based on SSEs as it is currently done for ionic-liquid gates.\\

Here we demonstrate a second generation of Li-ion conducting glass-ceramic (LICGC)-gated transistors based on mono/few layer semiconducting TMDs, whose performance rivals in all regards that of state-of-the-art ionic-liquid gated devices. This result was reached after having identified and eliminated the phenomena preventing the operation of LICGC gates at large electron densities. Specifically, the detailed comparison of existing LICGC and ionic-liquid gated devices, together with cyclic voltammetry (CV) experiments, reveal that difficulties encountered in past work originate from the production of redox species caused by faradic reactions at the interface between the LICGC substrate and the metal electrodes. We discuss how the redox reactions affect transistor operation and show that all parasitic effects can be eliminated by a SiO$_2$ passivation layer (40 nm) inserted between the LICGC and the metal electrodes. We perform Hall-effect measurements to extract the geometrical capacitance of the resulting devices and find it to be $C_\mathrm{G}\approx 20 \:\mathrm{\mu F/cm^2}$ and $C_\mathrm{G}\approx 50 \:\mathrm{\mu F/cm^2}$ for hole and electron accumulation, respectively, comparable to that of ionic liquids\cite{IL_Review_2017,CrossQuantum_2019}. The level of control demonstrated here enables the correct implementation of ionic-gate spectroscopy and the accumulation of carrier densities $> 1\times 10^{14}\:\mathrm{cm^{-2}}$. We further show that at these electron densities LICGC-gated devices exhibit superconductivity below a few Kelvin. These results demonstrate the reliable and reproducible operation of transistors enabling high density electrostatic gating with solid-state electrolytes.

\section{Results and Discussion}
We select as solid-state electrolyte LICGC (see Fig. 1a-c) substrates of the NASICON type\cite{goodenough1976fast,knauth2009inorganic,manthiram2017lithium}, whose chemical compositions are Li$_2$O-Al$_2$O$_3$-SiO$_2$-P$_2$O$_5$-TiO$_2$-GeO$_2$ (AG-01; purchased from Ohara Corporation\cite{Ohara}) and Li$_2$O-Al$_2$O$_3$-SiO$_2$-P$_2$O$_5$-TiO$_2$ (LASPT; purchased from MTI Corporation but synthesized by Ohara\cite{Ohara}). Produced by drawing of the melted raw materials, these glass-ceramics contain an oxide matrix where Li$^+$ ions can freely move through interstitial sites, resulting in room-temperature ionic conductivities (1-4$\:\times\: 10^{-4}\:\mathrm{S/cm^2}$\cite{Ohara}) comparable to those of liquid electrolytes\cite{Frisbie_review_2013,DEME-TFSI_Condivity}. Additionally, the LICGCs are chemically stable under ambient conditions, have a large electrochemical window (3 V, $-3$ V) and a sufficiently low surface roughness (a root mean square $\mathrm{Rms}=1\:\mathrm{nm}$ calculated from the topography image Fig 1c), which ensures a sufficiently intimate contact between the glass-ceramic and the gated material (other commercially available glass-ceramics that we tested had significantly larger surface roughness, which made working with atomically thin 2D materials impossible).\\

\begin{figure*}[t]
  \includegraphics[width=\linewidth]{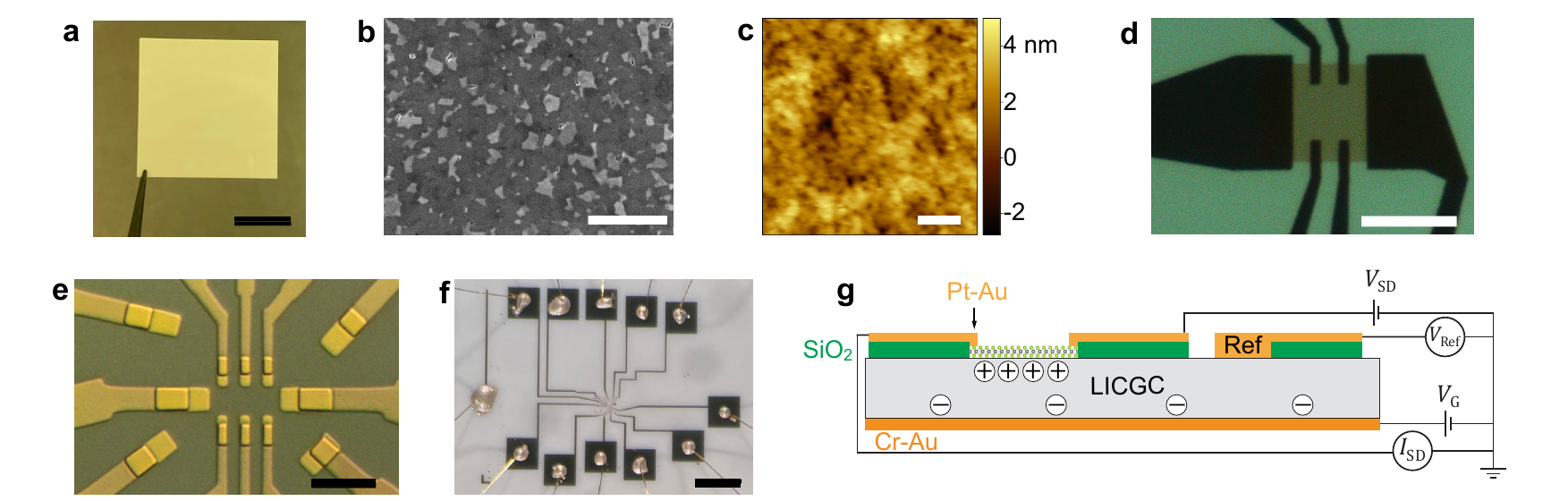}
  \caption{Li-ion conducting glass-ceramic gated FETs. \textbf{a)} Optical image of an as-purchased LICGC substrate; the scale bar is 1 cm. \textbf{b)} Scanning electron microscope image of the surface of the LICGC. The grains of different sizes (hundreds of nanometers) correspond to the different oxides that make up the glass-ceramic\cite{Ohara,katzenmeier2021properties}. The scale bar is $1\,\mu\mathrm{m}$. \textbf{c)} Contact-mode atomic force topography image of the surface of a LICGC substrate, from which we calculate a 1 nm rms roughness; the scale bar is $2\,\mu\mathrm{m}$. \textbf{d)} Optical microscope images of a 4L-WSe$_2$ LICGC-gated FET contacted with metal electrodes in direct contact with the LICGC substrate (image taken under a cross-polarization detection scheme, as described in REF.\cite{Polarizer_2021}).  \textbf{e)} Optical microscope image (without corss-polarization detection) of a device with a 40 nm SiO$_2$ layer between the metal electrodes and the LICGC (the overlay between the part of the electrodes where SiO$_2$ is present and absent is clearly visible). The scale bar in d) and e) is $10\,\mu\mathrm{m}$. \textbf{f)} Optical microscope image of a completed device, also showing the wires connecting the pad to the chip carrier used for electrical measurements. The scale bar is 1 mm. \textbf{g)} Schematic side view of a LICGC-gated FET with a SiO$_2$ passivation layer  (green layer in the schematics), showing the Pt contacts, the reference (Ref) electrode and the Cr-Au backgate.}
  \label{fig:1}
\end{figure*}

To illustrate the problems affecting LICGC back-gated FETs, we compare the behavior of such devices to that of conventional top-gate ionic-liquid gated transistors (top-gated FETs based on other commonly used electrolytes also provide an adequate means of comparison, and leads to identical conclusions). The LICGC FETs (see Fig. 1d-g) are realized employing mono/few layer WSe$_2$ and MoS$_2$ exfoliated from bulk crystals (purchased from HQ Graphene). The layers are exfoliated onto Si/SiO$_2$ substrates and subsequently transferred by a conventional pick-up and release technique\cite{zomer2014fast} onto the LIGCG substrates whose back side is coated with an evaporated Cr/Au (10/70 nm) layer acting as gate electrode. Contacts to the WSe$_2$ and MoS$_2$ layer are fabricated by means of electron beam lithography, evaporation of a Pt/Au layer (5/45 nm), and lift-off. Depending on the device, the evaporated Pt is either in direct contact with both the substrate and the WSe$_2$ and MoS$_2$ layer, or only with the WSe$_2$ and MoS$_2$ layer. In the latter case we deposit a 40 nm SiO$_2$ layer (capped with Ti/Au\cite{SiO2_2009} (5/25 nm)) between the Pt and the LICGC substrate, as schematically shown in Fig. 1g (see experimental section for more details). A reference electrode\cite{RefElectrode_2009,ILS_review_2021} (see Fig. 1e) is also present: in properly functioning devices, the potential $V_\mathrm{Ref}$ measured between the reference and the source electrode corresponds to the voltage drop across the electrolyte/channel interface.  $V_\mathrm{Ref}$ serves to determine the “gate efficiency” (defined as $\Delta V_\mathrm{Ref}/\Delta V_\mathrm{G}$), a quantity that measures how effectively the applied potential is transferred from the gate to the transistor channel. To maximize the gate efficiency, the area between the gate electrode and the electrolyte (and hence the capacitance) is intentionally designed to be much larger than the electrolyte/device area, which includes the semiconductor and the electrodes \cite{ILS_review_2021,wong2017impact,joshi2018understanding}. The ionic-liquid gated FETs that we studied for comparison are realized as repeatedly detailed in our earlier works\cite{Spectroscopy_2012,LET_2014_M,artificial_2018,MoS2_2018,CrossQuantum_2019,ILS_review_2021}: WSe$_2$ is exfoliated onto a Si/SiO$_2$ substrate, Pt/Au (5/45 nm) contacts are attached with conventional nano-fabrication techniques, and an ionic liquid (DEME-TFSI) is drop-casted as a final step. Unless otherwise specified, all the measurements are performed at room temperature.\\

%, i.e., the amount of gate voltage that drops across the electrolyte/device interface, the area between the gate electrode and the electrolyte (and hence the capacitance) is intentionally designed to be much larger than the electrolyte/device area (including the semiconductor and the electrodes) \cite{ILS_review_2021}.
\begin{figure*}[t]
  \includegraphics[width=\linewidth]{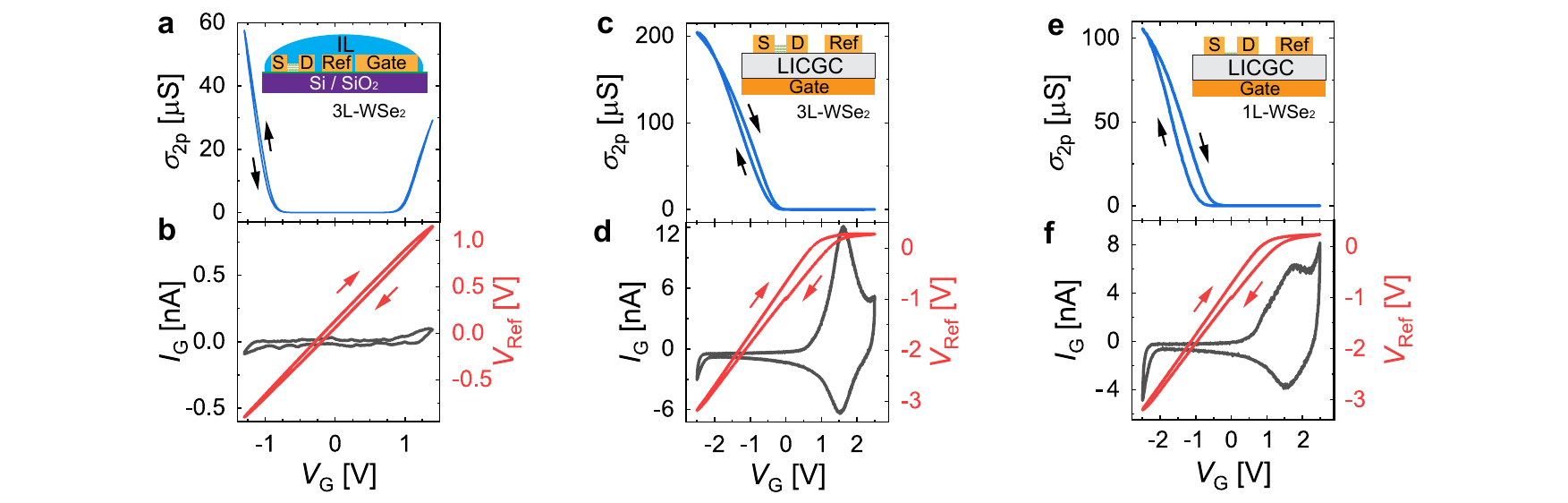}
  \caption{Comparison of ionic-liquid and LICGC gated FETs. \textbf{a)} Two-terminal conductivity $\sigma_\mathrm{2p}$ vs gate voltage $V_\mathrm{G}$ (i.e., transfer curves) measured on a 3L-WSe$_2$ ionic-liquid gated FET (see inset for a device schematics), showing ideal ambipolar characteristics. \textbf{b)} On the same device the gate current (black curve) is small and featureless, and the reference potential (red curve) varies linearly with gate voltage. Analogous measurements are shown in panels \textbf{c)}-\textbf{d)} and \textbf{e)}-\textbf{f)} for a 3L- and a 1L-WSe$_2$ LICGC-gated FET, respectively (see the insets in panels c) and e) for device schematics). For positive gate voltage, no transport is observed, the leakage current $I_\mathrm{G}$ exhibits large peaks signaling the presence of chemical reactions, and the reference potentials saturate (the $V_\mathrm{G}$ sweep rate is 3 mV/s in a)-b) and e)-f), and 2 mV/s in c) and d); the different magnitude of the $I_\mathrm{G}$ peaks in d) and f) are caused by the different area of the devices, whose largest contribution is the Pt/LICGC interface, and not the different sweep rates.}
  \label{fig:2}
\end{figure*}

%\textcolor{red}{}
%\textcolor{blue}{\sout{}}

The difference in behavior of a “conventional” ionic-liquid gated transistor and of a LICGC-gated device is obvious from the data in Fig. 2, which compares measurements performed on the two types of devices. The transfer curve (conductivity $\sigma_{\mathrm{2p}}$ versus  $V_\mathrm{G}$) of ionic-liquid devices (Fig. 2a) exhibit balanced ambipolar transport, resulting in relatively high electron and hole conductivities as $V_\mathrm{G}$ exceeds the respective threshold voltage ($V\mathrm{_{Th}^h}=-0.9\:\mathrm{V}$ and $V\mathrm{_{Th}^e}=0.9\:\mathrm{V}$ in this device). The reference potential $V_\mathrm{Ref}$ (Fig. 2b) varies linearly with $V_\mathrm{G}$ and the gate efficiency is typically 80-90\% (80\% in this device). The current measured at the gate electrode ($I_\mathrm{G}$) is small and of capacitive nature. These features are characteristic of high-quality ionic-liquid gated devices\cite{Spectroscopy_2012,LET_2014_M,MoTe2_2014,WS2_SC_2015,ReS2_2016,MoS2_SC_2016,MoS2_2018,costanzo2018tunnelling,CrossQuantum_2019}, such as the ones that we commonly employ to perform ionic gate spectroscopy \cite{Spectroscopy_2012,MoTe2_2014,MoS2_2018,ReS2_2016,CrossQuantum_2019,ILS_review_2021}, and to study gate-induced superconductivity (when materials such as MoS$_2$ and WS$_2$ are used)\cite{WS2_SC_2015,MoS2_SC_2016,costanzo2018tunnelling}. \\

The behavior is different for LICGC-gated FETs. Although transistor action is seen for hole accumulation ($V_\mathrm{G}<0\:\mathrm{V}$ in Figs. 2c and e) and the same conductivity is present in the LICGC and ionic-liquid gated devices at similar $\Delta V\mathrm{_G^h}=V_\mathrm{G}-V\mathrm{_{Th}^h}$ values ($\approx$ 50 $\mu$S at $\Delta V\mathrm{_G^h}=0.5\:\mathrm{V}$; compare Fig. 2a to Figs. 2c and e), there is no transistor action for electron accumulation. Indeed, for $V_\mathrm{G}>0\:\mathrm{V}$ the conductivity remains vanishingly small (in some device a slight increase is observed, followed by saturation at small positive $V_\mathrm{G}$). $V_\mathrm{Ref}$ depends linearly on $V_\mathrm{G}$ for negative applied gate voltage, and saturates for $V_\mathrm{G}>0.75\:\mathrm{V}$ (Fig. 2d), indicating that under these conditions the applied gate voltage is not transferred to the interface with the semiconductor. Additionally, for $V_\mathrm{G}>0\:\mathrm{V}$ the current $I_\mathrm{G}$ measured at the back-gate electrode exhibits a non-monotonic dependence on $V_\mathrm{G}$. Clearly the observed behavior indicates that for positive gate voltage the devices do not operate as expected for field-effect transistors.\\

%\textcolor{blue}{\sout{Clearly the observed behavior indicates}} \textcolor{red}{These spurious effects are present in  1L- (Figs. 2c and d), 3L- (Figs. 2e and f) and bulk-WSe$_2$\cite{Marc_LICGC} LICGC-gated FETs, indicating} that for positive gate voltage the device\textcolor{red}{s} \textcolor{blue}{\sout{does not}} \textcolor{red}{don't} operate as expected for a field-effect transistor \textcolor{red}{independently of the thickness of the gated material.}
 \begin{figure*}[t]
  \includegraphics[width=\linewidth]{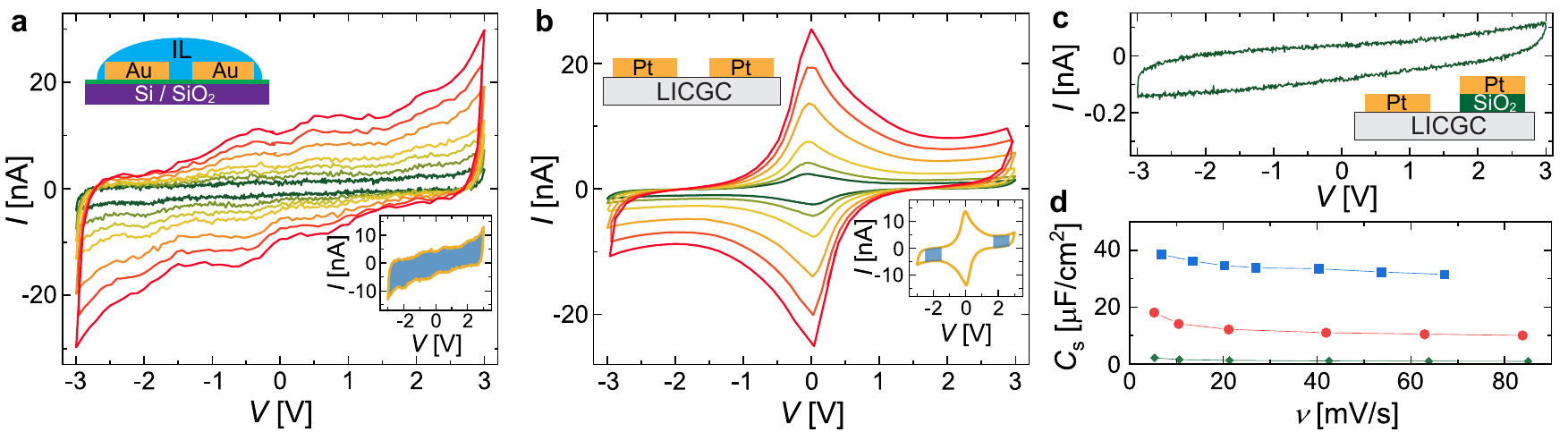}
  \caption{Cyclic voltammetry measurements with ionic liquids and LICGC substrates. \textbf{a)} CV curves measured in a device where the ionic liquid covers two Au electrodes (from the green to the red curve the sweep rates are 7, 13, 20, 27, 40, 54 and 67 mV/s). \textbf{b)} CV curves measured in a device where two Pt electrodes are deposited onto the LICGC (from the green to the red curve the sweep rates are 5, 10, 21, 42, 63 and 84 mV/s). The top insets in a) and b) show the schematic of the devices and the bottom insets the area of the curves used in the calculation of the capacitance shown in d). \textbf{c)} A CV curve measured in a device where a SiO$_2$ passivation layer is placed between one of the Pt electrode and the LICGC, as shown schematically in the inset (the sweep rate is 5 mV/s; see Fig. S2a for CV curves measured at different sweep rates). \textbf{d)} Geometrical capacitance per unit area $C_\mathrm{S}$ as a function of sweep rate extracted from the CV curves shown in a) (blue curve), b) (red curve) and Fig. S2a (green curve). The values represent the total capacitance resulting from the in-series addition of the capacitances at the two electrodes.}
  \label{fig:3}
\end{figure*}

Besides illustrating the parasitic phenomena affecting LICGC gated devices, the measurements in Fig. 2d provide an indication as to the origin of the problem, because the features observed in the gate current $I_\mathrm{G}$ are reminiscent of $I$-$V$ characteristics measured in Li-ion batteries\cite{CV_review_2018,ECtextbook_Bard}. Given that for $V_\mathrm{G}>0\:\mathrm{V}$ a large concentration of reactive Li$^+$ ions are accumulated at the interface with the active part of the transistors, it seems likely that interfacial electrochemical reactions involving these ions indeed occur in our devices. The measurements in Fig. 2d are not enough to determine whether the reactions take place at the interface with the semiconducting 2D material, the metal electrodes, or both, which is important to understand their effect on the device. It can nevertheless be concluded that they do not irreversibly compromise the integrity of the devices, simply because the same behavior is observed if the sweeps of applied gate voltage are repeated continuously over a period of many days (see supporting information). \\

To determine whether the chemical reactions involve the metal or the semiconductor part of our transistors, we realized two-terminal devices consisting of two metal electrodes in contact with either an ionic liquid or the LICGC (see the upper insets of Fig. 3a and 3b for a scheme of the devices). In these devices, sweeping the voltage applied between the two electrodes and measuring the current corresponds to performing CV measurements\cite{elgrishi2018practical,CV_review_2018,sandford2019synthetic,CV_review_2019,ECtextbook_Bard}. Measurements performed for different sweeping rates are shown in Fig. 3a and b for the ionic liquid and the LICGC, respectively. The current across the Au/ionic-liquid/Au device (referred to as CV-IL) traces a virtually featureless, quasi-rectangular hysteresis loop --whose magnitude increases with increasing sweeping rate-- as the applied voltage is swept from -3V to 3V (anodic sweep) and back (cathodic sweep). The current is also seen to depend linearly on sweep rate and extrapolates to 0 $\mathrm{A}$ upon decreasing the sweep rate (Fig. S1d). These observations are all consistent with a displacement current charging an EDL at the ionic-liquid/Au interface\cite{Frisbie_review_2013,Geo-IL_2016}. \\

In contrast, in the Pt/LICGC/Pt device (referred to as CV-GC) a pronounced peak in current centered at $V\approx 0\:\mathrm{V}$ is observed for both sweep directions, which adds onto the displacement current visible at larger positive and negative voltage. In CV measurements performed on electrochemical cells containing Li$^+$ ions these peaks typically originate from the formation of coupled redox species caused by faradaic reactions of the type $x\mathrm{Li}^++x\mathrm{e}^-+\mathrm{R}\rightleftharpoons\mathrm{Li}_x\mathrm{R}$ (R is the reduced species), taking place at the electrolyte/electrode interface during the anodic and cathodic sweep, commonly resulting in the formation of a self-limiting Nernstian diffusion layer\cite{elgrishi2018practical,CV_review_2018,famprikis2019fundamentals,ECtextbook_Bard}. In the anodic sweep the current increases initially due to the exchange of electrons at the interface and --because of the diffusion layer--  subsequently decreases leading to the observed peak, as the product formed during the reaction hinders more reactants from reaching the interface (the absolute value of the peak current increases with sweeping rate because a larger sweeping rate results in a thinner Nernstian diffusion layer\cite{ECtextbook_Bard}). The same holds true for the cathodic sweep, with the reaction occurring in the reverse direction. At the LICGC/Pt interface of our devices, the most likely redox reaction taking place is the reduction of Ti in the Li$_x$TiO$_2$\cite{LiTiO_1997,LiTiO_2007} present in the solid electrolyte, as it has been reported for a similar LICGC\cite{amiki2013electrochemical,TiLi_2017}, with details that depend on the specific metal electrode used
(as shown in the supporting information, CV measurements performed on a Cr/LICGC/Cr device several peaks are present, and not only one as for Pt/LICGC/Pt devices). \\

The occurrence of electrochemical reactions is consistent with the different voltages at which the current peaks  (electrochemical cell potential) are observed in the CV-GC device just discussed ($\approx 0\:\mathrm{V}$; Fig. 3b) and in the LICGC-gated FET shown Fig. 2d ($\approx 1.5\:\mathrm{V}$; in this case, the relevant current is that measured at the gate electrode). That is because the two devices effectively correspond to different types of electrochemical cells. Specifically, the CV-GC device (with equal electrode areas) can be considered as a full cell\cite{ECtextbook_Bard}, where the electrochemical cell potential is $\approx 0\:\mathrm{V}$ because the same redox reaction occurs at the anode and the cathode. Instead, the gate-electrode/LICGC/Pt structure of the FET behaves as a half cell\cite{elgrishi2018practical,ECtextbook_Bard} because it is intentionally designed to maximize the voltage drop across the LICGC/Pt interface (i.e., to maximize the gate efficiency, see above). In the latter case, the cell potential ($\approx 1.5\:\mathrm{V}$) is therefore solely determined by the reaction taking place at the LICGC/Pt interface created to establish electrical contact to the 2D semiconductor. \\

Irrespective of the nature of the redox reactions (whose identification is not the purpose of this work) what is essential is to understand how their presence hinders the correct operation of the LICGC-gated FETs (Fig. 2c and d), and whether the unwanted effects that they cause can be eliminated. Because of the reaction of Li$^+$ ions forming neutral species at the he LICGC/Pt interface, any further increase of $V_\mathrm{G}$ past the cell potential (at $\approx 0.75 \:\mathrm{V}$; Fig. 2d) does not result in the accumulation of additional electrons and Li$^+$ ions. The density of accumulated charge on each side of the interface remains then constant, and so does the electric field, as signaled by the saturation of $V_\mathrm{Ref}$ (see Fig. 2d). Since no additional potential is transferred to the device interface as  $V_\mathrm{G}$ is increased, the gate looses its ability to control the chemical potential in the semiconductor, which is why the transistor does not function properly. In the simplest possible terms, the type of process that causes problems for FET operation is the same that makes batteries work:  electrochemical reactions allow the applied gate voltage to store energy through an increased density of neutral chemical species, and not in the form of electrostatic energy\cite{conway1991transition} (i.e., by increasing the electric field and charge density). An additional problem for transistor operation is that the $I_\mathrm{G}$ peaks created by Li$^+$ ions involved in the redox reaction creates a ohmic voltage drop in the interior of the glass-ceramic, reducing even more the  potential transferred to the semiconductor surface. These considerations suggest that correct FET operation may be possible when a passivation layer or a passivation layer is placed at the LICGC/Pt interface to separate the species involved in the reaction (i.e., TiO$_2$, Li$^+$ ions and electrons. We tested the idea by performing CV measurements on a CV-GC device in which a SiO$_2$ layer was inserted between the glass-ceramic and one of the Pt electrodes (see inset of Fig. 3c). The absence of any redox peaks in Fig. 3c indicates that the insertion of SiO$_2$ eliminates all manifestations of redox reactions (and reduces the current due to its smaller capacitance; compare Fig. 3a and b with c). Since Li$^+$ ions likley migrate into the SiO$_2$ layer, we believe its function is to spatially separates the TiO$_2$ present in the ceramic from the electrons in the Pt contact, preventing the formation of Li$_x$TiO$_2$ (see supporting information for more details).\\

\begin{figure}[t]
  \includegraphics[width=\linewidth]{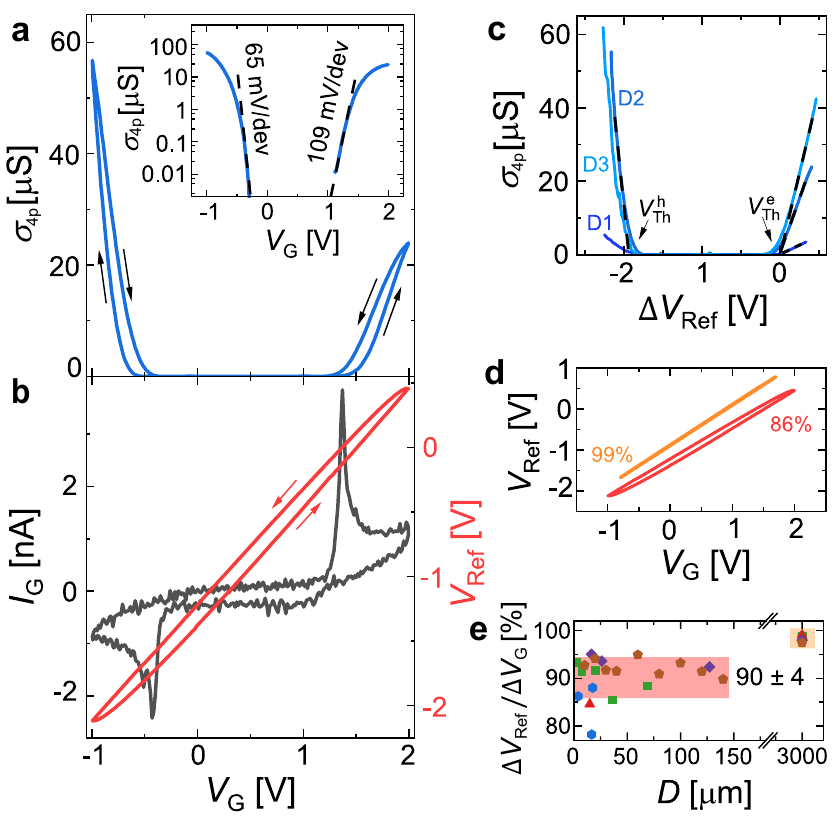}
  \caption{High-quality ambipolar LICGC-gated 1L-WSe$_2$ FETs, with a SiO$_2$ passivation layer  between the electrodes and the LICGC substrate (in all measurements the sweep rate is below $10\,\mathrm{mV/s}$). \textbf{a)} The four-probe conductivity $\sigma_\mathrm{4p}$ vs $V_\mathrm{G}$ exhibits high quality ambipolar behavior. Inset: same plot in semi-log scale; the dashed black lines are used to extract the slope of the curve to obtain the sub-threshold swings. \textbf{b)} Reference potential $V_\mathrm{Ref}$ (red curve) and leakage current $I_\mathrm{G}$  (black curve) measured as a function of $V_\mathrm{G}$ on the same device whose transfer curve is shown in a). The small peaks in the leakage current indicate the onset of electron and hole accumulation (i.e., are not due to chemical reactions). \textbf{c)} Ambipolar transfer curves of three different devices (D1, D2 and D3), as a function of $\Delta V_\mathrm{Ref}$ = $V_\mathrm{Ref}$ $-$ $V_\mathrm{Th}^\mathrm{e}$. The dashed lines illustrate the determination of the threshold voltages by the linear extrapolation of $\sigma_{4p}$ to $\sigma_{4p}$ = 0 S, performed to determine the bandgap as $\Delta$ = $e(V_\mathrm{Th}^\mathrm{e}$ $-$ $V_\mathrm{Th}^\mathrm{h})$. \textbf{d)} Potential measured in device D2 by a reference electrode several millimeters away from the device (orange curve) and by a reference electrode positioned within approximately 100 microns or less from the device (red curve); the far away electrodes always exhibit nearly 100\% efficiency ($\Delta V_\mathrm{Ref}/\Delta V_\mathrm{G}$), whereas the nearby electrode varies typically between 85 and 95\%, as summarized in panel  \textbf{e)} that includes measurements on many different devices and reference electrodes.}
  \label{fig:4}
\end{figure}

Before testing LICGC-gated transistors with a SiO$_2$ passivation layer preventing direct contact between the metal and the glass-ceramic, we analyze the CV measurements on the CV-IL and CV-GC devices, to estimate their geometrical capacitance. We do this by calculating the area enclosed by the anodic and cathodic sweeps shown in Fig. 3a and 3b, as commonly done in the literature\cite{yu2014scalable,li2018new,kim2020applications} (for LICGC, we take care to calculate the area by selecting a voltage interval not affected by the peaks originating from the redox reactions). The resulting total capacitance per unit area $C_\mathrm{S}$ as a function of sweeping rate is plotted in Fig. 3d. $C_\mathrm{S}$ is given by the in-series connection of capacitances of the EDLs at the two metal/electrolyte interfaces, i.e.,  $1/C_\mathrm{S}=1/C_\mathrm{EDL1}+1/C_\mathrm{EDL2}$, and is dominated by the smallest of the two EDL capacitances. For ionic liquid, the smaller geometrical capacitance is the one for hole accumulation (when anions are present at the device interface), whose known value\cite{CrossQuantum_2019} is in very good agreement with the value extracted from Fig. 3d ($40\: \mathrm{\mu F/cm^2}$ at the lowest sweep rate). For the LICGCs, we find $C_\mathrm{S} = 20\: \mathrm{\mu F/cm^2}$ slightly smaller, but comparable to that of ionic liquids\cite{IL_Review_2017,CrossQuantum_2019}. More precise information about the geometrical capacitance of LICGC electrolytes is extracted below from Hall effect measurements performed on transistor devices. \\

%\textcolor{red}{}
%\textcolor{blue}{\sout{}}
%$\sigma_\mathrm{4p}$

The results of measurements performed on a 1L-WSe$_2$ LICGC-gated FET with a SiO$_2$ passivation layer separating the metal contacts and the LICGC are shown in Fig. 4a and b. The device behavior differs drastically from that of the WSe$_2$ FETs without passivation layer (Figs. 2c-f). In particular, well-balanced ambipolar transport is observed, with electron conductivity values as high as in ionic-liquid FETs ($\approx$ 30 $\mu$S at $\Delta V\mathrm{_G^e}=V_\mathrm{G}-V\mathrm{^e_{Th}}=0.5\:\mathrm{V}$; compare Fig. 4a to Fig. 2a). $V_\mathrm{Ref}$ follows a linear dependence on $V_\mathrm{G}$ throughout the entire applied $V_\mathrm{G}$ range (Fig. 4b) with an efficiency of $\approx86\%$ (comparable to that of ionic-liquid gated devices; Fig. 2b), with no sign of saturation, and of chemical reactions in the current $I_\mathrm{G}$ measured at the gate electrode (the sharp peaks in $I_\mathrm{G}$ coincide with the onset of hole and electron conduction and are a manifestation of the onset of charge accumulation in the transistor channel\cite{Spectroscopy_2012}; see Fig 4a). The high geometrical capacitance results in low sub-threshold swings (see inset of Fig. 4a), approaching the ultimate room temperature limit of $\approx 60\:\mathrm{mV/dec}$\cite{sze_physics_2006}. In short, devices with a SiO$_2$ passivation layer exhibit high-quality transistor characteristics, with no indications of the effect of redox reactions (from which we conclude that the reactions do not involve WSe$_2$) As we proceed to demonstrate, LICGC-gated FETs enable the realization of experiments normally performed with high-quality ionic-gated FETs, such as ionic-gate spectroscopy and gate-induced superconductivity (by safely reaching gate voltages beyond those shown in Fig. 4.)\\

We start by demonstrating the ability to perform ionic-gate spectroscopy with LICGC-gated FETs. As we established in our past work on ionic-liquid gated devices\cite{Spectroscopy_2012,MoTe2_2014,MoS2_2018,ReS2_2016,CrossQuantum_2019,ILS_review_2021}, the gap of semiconductors can be extracted directly from the transfer curve of transistors having a sufficiently large geometrical capacitance, as $\Delta=e(V\mathrm{^e_{Th}}-V\mathrm{^h_{Th}})$ (where $V\mathrm{^e_{Th}}$ and $V\mathrm{^h_{Th}}$ are the threshold voltage for electron and hole conduction respectively; for a comprehensive review on ionic-gate spectroscopy see Ref. \cite{ILS_review_2021}). Fig. 4c shows the ambipolar transfer curves (four-probe conductivity $\sigma_\mathrm{4p}$ vs $V_\mathrm{Ref}$) of three different LICGC-gated FETs (D1-D3) measured to quantify the band gap of 1L-WSe$_2$ (for ease of comparison, the curves are plotted by taking in each device $V\mathrm{^e_{Th}}$ as reference for the voltage). The voltage range in which no current flows --which measures the magnitude of the gap-- is nearly identical in all cases, resulting in a value of the gap of 1L-WSe$_2$ $\Delta_\mathrm{1LWSe_2}=1.93\pm 0.07 \:\mathrm{eV}$ (the error is less than 5\%, and is determined by the spread in the measurements on the different devices), only slightly larger than the value $\Delta_\mathrm{1LWSe_2}=1.81\pm 0.06 \:\mathrm{eV}$ extracted using ionic-liquid gated devices\cite{ILS_review_2021}. The difference is possibly due to the different dielectric constants of IL and LICGC and the strain induced by the substrate, which are known to affect the precise value of the band gap\cite{Strain_2012,chernikov2014exciton,peng2020strain,riis2020electrically,park2021schottky}. We have also performed ionic-gate spectroscopy on 2L-WSe$_2$ (see supporting information) and obtained $\Delta_\mathrm{2LWSe_2}=1.63\pm 0.06 \:\mathrm{eV}$, in good agreement with the $1.61\pm 0.05 \:\mathrm{eV}$ value found using ionic-liquid gates\cite{ILS_review_2021}. We note that the ability to obtain precise values of the gap using LICGC gates relies on the implementation of reference electrodes positioned within $150\:\mathrm{\mu m}$ of the channel of the device (see experimental section for details on their fabrication). Indeed, we found that employing large reference electrodes positioned millimeters away from the device (see Fig. 1f) results in an overestimation of $V_\mathrm{Ref}$ (see Figs. 4d and 2) and consequently in gap values larger than the actual one, an effect  absent in ionic-liquid gated FETs and whose origin remains to be understood. The reason for this difference is possibly that in ionic liquid devices the reference electrodes are perfectly coupled electrostatically to the electrolyte (as they are covered by it), which screens spurious electrostatic potentials from other sources, whereas for LICGC gates the reference electrodes are on the surface of the solid-state electrolyte, and therefore not fully screened from outside potentials. \\

%\textcolor{red}{}
%\textcolor{blue}{\sout{}}
\begin{figure}[t]
  \includegraphics[width=\linewidth]{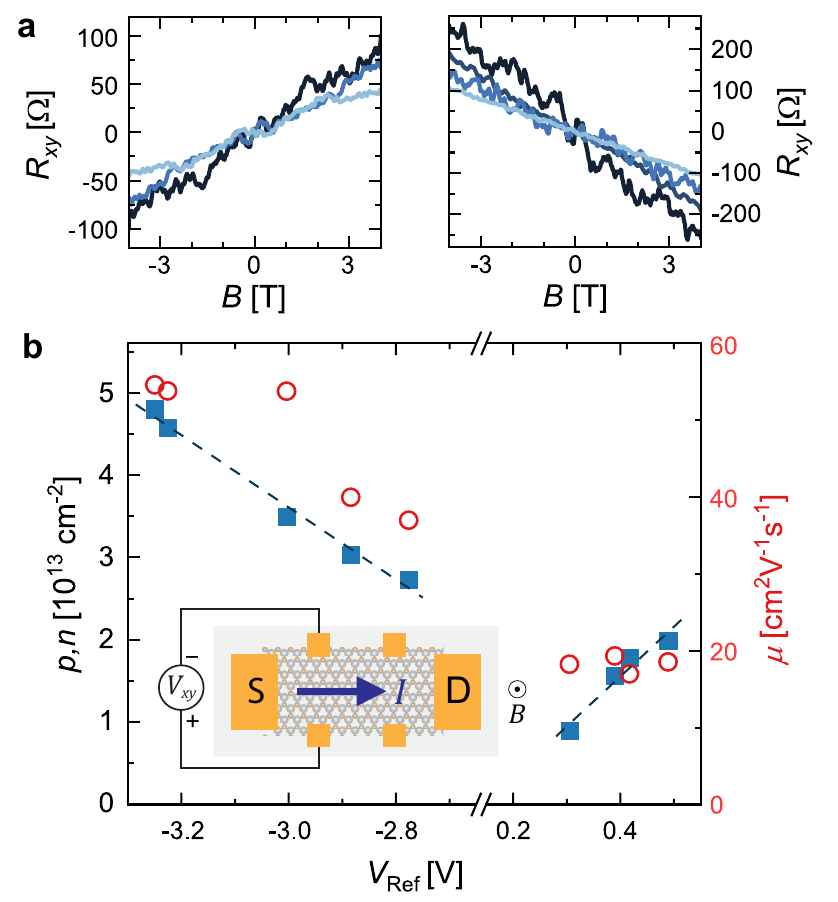}
  \caption{Hall measurements performed on a 1L-WSe$_2$ LICGC-gated FET. \textbf{a)} Magnetic field $B$ dependence of the (antisymmetrized) transversal resistance $R_{xy}$ measured at different gate voltages, under hole (left panel) or electron (right panel)  accumulation (for holes $V_\mathrm{Ref}$ = $-2.88$, $-3$ and $-3.25$ V and for electrons $V_\mathrm{Ref}$ = 0.31, 0.39, 0.42, 0.49 V, as the curve color progresses from darker to lighter blue). \textbf{b)} Carrier density (blue squares) and mobility (red open circles) extracted from the Hall measurements, plotted as a function of $V_\mathrm{Ref}$. The dotted blue lines represent linear regressions performed to extract the capacitance of the device (the inset shows the measurement configuration used to extract the Hall resistance).}
  \label{fig:5}
\end{figure}

To further characterize the LICGC-gated devices we measured their capacitance, by performing Hall effect measurements under electron and hole accumulation on one of our 1L-WSe$_2$ LICGC-gated devices (D3). The  transverse resistance $R_{xy}$ as a function of applied magnetic field $B$ measured for different reference potentials is shown in Fig. 5a, and the resulting  hole and electron densities are plotted in Fig. 5b. The values are comparable to those measured on ionic-liquid gated devices for comparable values of applied gate voltage\cite{MoTe2_2014,CrossQuantum_2019}. Since the total device capacitance is given by $1/C_\mathrm{T}=1/C_\mathrm{G}+1/C_\mathrm{Q}$ (where $C_\mathrm{Q}$ is the quantum capacitance associated of 1L-WSe$_2$), the  measured electron and hole densities allow us to extract the geometrical capacitance of the devices. To obtain the value of $C_\mathrm{G}$, we first extract $C_\mathrm{T}=e\mathrm{d}n/\mathrm{d}V_\mathrm{Ref}$ by performing a linear regression of $n(V_\mathrm{Ref})$ and $p(V_\mathrm{Ref})$ (dashed lines in Fig. 5b), resulting in values of $\approx 7 \:\mathrm{\mu F/cm^2}$ and $\approx 10 \:\mathrm{\mu F/cm^2}$ for hole and electron accumulation, respectively. We then, use the experimental known values of $C_\mathrm{Q}$ for 1L-WSe$_2$ (including the effect of the cross-quantum capacitance\cite{CrossQuantum_2019,CrossQuantum_2021}; $\approx 13 \:\mathrm{\mu F/cm^2}$ for holes in the first spin-split K valley of the valence band and $\approx 12 \:\mathrm{\mu F/cm^2}$ for electrons in the two K valleys of the conduction band\cite{CrossQuantum_2019}), and find that for LICGC gates $C_\mathrm{G}\approx 16 \:\mathrm{\mu F/cm^2}$ and $\approx 50 \:\mathrm{\mu F/cm^2}$ for hole and electron accumulation, only slightly smaller than the geometrical capacitances of ionic-liquid gates (respectively $C_\mathrm{G}\approx 40 \:\mathrm{\mu F/cm^2}$ and $\approx 60 \:\mathrm{\mu F/cm^2}$\cite{CrossQuantum_2019} for hole and electron accumulation). Note that for LICGC connecting in series the hole and electron geometrical capacitances gives a value of $\approx 12 \:\mathrm{\mu F/cm^2}$, close to the value estimated from the measurements performed on CV-GC devices discussed earlier, demonstrating the internal consistency of our results).\\

\begin{figure}[t]
  \includegraphics[width=\linewidth]{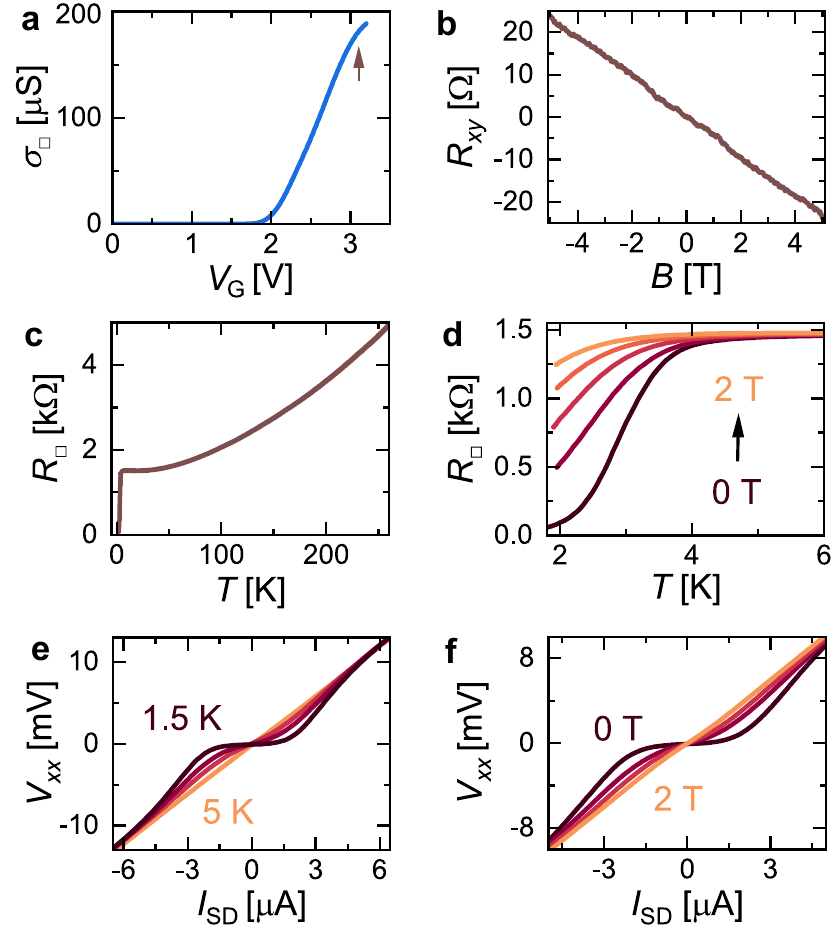}
  \caption{Gate-induced superconductivity in a 3L-MoS$_2$ LICGC-gated FET. \textbf{a)} Four-probe conductivity $\sigma_\mathrm{4p}$ as a function of gate voltage $V_\mathrm{G}$. \textbf{b)} Magnetic field $B$ dependence of the (antisymmetrized) transversal resistance $R_{xy}$ at $V_\mathrm{G}$ = 3.1 V. \textbf{c)} Temperature $T$ dependence of the four-probe square resistance $R_\square$ measured at $V_\mathrm{G}$ = 3.1 V, showing a sharp drop below 4 K, indicating the presence of gate-induced superconductivity. \textbf{d)} $R_\square$ measured at different out-of-plane magnetic fields $B$ for temperatures close to the superconducting transition. \textbf{e)} and \textbf{f)} show the dependence of the superconducting critical current on $T$ at $B$ = 0 T (from darker to lighter color, $T$ = 1.5, 2.5, 3 and 5 K) and $B$ at $T$ = 1.7 K (from darker to lighter color, $B$ = 0, 0.5, 1 and 2 T).}
  \label{fig:6}
\end{figure}

Finally, we search for the the occurrence of gate-induced superconductivity in the presence of sufficiently large electron density in MoS$_2$\cite{MoS2_SC_2012,TMDSC_2015,MoS2_SC_2016,costanzo2018tunnelling,piatti2018multi}. To this end, low-temperature transport measurements were performed on a 3L-MoS$_2$ LICGC-gated FET.  Before cooling down the device, the gate voltage was swept to $V_\mathrm{G} = 3.1\:\mathrm{V}$ (see Fig. 6a) at room temperature, with Hall measurements (Fig. 6b) revealing an electron density of $1.25\times 10^{14}\:\mathrm{cm^{-2}}$ (in very good agreement with the density accumulated with ionic-liquid gates at similar values of $\Delta V\mathrm{_G^e}$, confirming that charge accumulation is of electrostatic nature). Fig. 6c shows the temperature dependence of the four-probe square resistance $R_\square$ measured as a function of temperature $T$. $R_\square$ exhibits a sharp drop below $\approx 4\:\mathrm{K}$ (see Fig. 6d), originating from the transition to the superconducting state at $T_\mathrm{C} = 2.9\:\mathrm{K}$ (corresponding to a drop of 50\% in $R_\square$). The $I$-$V$ characteristics of the device exhibit a sizable supercurrent that is suppressed with increasing temperature (see Fig. 6e) and by applying magnetic field (see Fig. 6f), and disappears above $T_\mathrm{C}$ and a critical field $B_\mathrm{C} \approx 0.9\:\mathrm{T}$. We conclude that LICGC-based FETs do enable the accumulation of electron density in excess of 10$^{14}$ cm$^{-2}$, resulting in gate-induced superconductivity, demonstrating that their quality remains comparable to that of more conventional ionic-liquid gated devices even at cryogenic temperatures.\\

\section{Conclusion}
A lesson learned from early days of research on ionic-gating  is that exploiting the full capability of electrolytes  relies on the identification and elimination of the unwanted electrochemical processes taking place in the devices. Early work, for instance, showed that traces of water and oxygen present in ionic liquids can reduce their electrochemical window\cite{IL_Water_2008,IL_Humidity_2020}, induce unwanted electrochemical reactions\cite{IL_O2_2004}, and even cause irreversible degradation of the gated material\cite{water_2012,ZnO_water_2012}. Similarly, it has been also understood that reversible electrostatic gating is very difficult to achieve for many transition-metal oxides\cite{jeong2013suppression,wang2020scattering}, because the large electric fields present in ionic-gated devices can pull oxygen atoms out of the materials. If not correctly identified, these processes lead to spurious effects and experimental artefacts that cause severe experimental irreproducibility, making the technique unreliable.  For ionic-liquid gating in combination with chemically inert materials such as graphene or semiconducting transition-metal dichalcogenides the situation is now fully under control. For other electrolytes of interest it is not, and --every time a new type of electrolyte is introduced-- research is needed to determine if the material can be operated controllably as ionic gates, and under which conditions.\\

It is in this context that the present work on Lithium-ion conducting glass-ceramic  for gating represent a significant advance in the domain of ionic gating. Recent experiments showing how promising these solid-state electrolytes are for gating were undoubtedly plagued by several parasitic effects of unknown origin, casting serious doubts about different aspects of the results, and about the reliability of the technique. By successfully identifying these spurious effects and eliminating them, our work makes now possible to reliably employ solid-state electrolytes in many different types of gating experiments. This will be particularly important for double ionic-gating experiments and for all experiments requiring gating to high carrier density (10$^{14}$ cm$^{-2}$ or higher), while keeping the material surface accessible to perform scanning tunneling microscopy or angle-resolved photo emission spectroscopy measurements. \\

\section{Experimental Section}
\textbf{LICGC-gated FETs}
LICGCs (AG-01 and LASPT) are purchased as $25\:\mathrm{mm}\times25\:\mathrm{mm}\times150 \:\mathrm{\mu m}$ substrates with polished top and bottom surfaces (resulting in a root mean square roughness of $1\:\mathrm{nm}$; see main text). Before cutting the substrates into $5\:\mathrm{mm}\times10\:\mathrm{mm}$ pieces their top surface is covered with a protective PMMA layer and their bottom surface with a Cr/Au (10/70 nm) layer deposited via ebeam-evaporation that will act as the gate electrode. Once the glass-ceramics are cut into smaller pieces, the PMMA is dissolved and the surface is cleaned with an oxygen plasma to remove any PMMA residue. The few-layer WSe$_2$ and MoS$_2$ crystals exfoliated onto SiO$_2$/Si substrates are then picked-up and transferred onto the top surface of the LICGCs using well-established procedures based on PC/PDMS stamps\cite{zomer2014fast}. On the ceramic substrates the few-layer crystals are invisible under normal microscope imaging conditions and a cross-polarized detection scheme is used to visualize them (see Fig. 1d and Ref.\cite{Polarizer_2021} for more details). The same detection scheme allows the identification of few-layer crystals via optical analysis, enabling the fabrication of devices using crystals exfoliated directly onto the LICGCs (we chose to transfer the crystals from SiO$_2$/Si substrates to avoid unnecessary consumption of ceramic substrates during the exfoliation process). The metal electrodes are fabricated directly onto the ceramic/crystals by means of standard nano-lithography techniques, as described in the main text. Two lithography steps are needed to fabricate the contacts and reference electrodes when a SiO$_2$ passivation layer is present between the Pt contacts and the LICGC. In the first step Al$_2$O$_3$/SiO$_2$/Ti/Au (2/40/5/25 nm) electrodes are fabricated until the edge of the 2D semiconductor crystals (Al$_2$O$_3$ is used as a sticking layer between SiO$_2$ and the TMD crystals, and the Ti/Au layer is used to cap the SiO$_2$ layer\cite{SiO2_2009} and provide a conducting path). In the second step a Pt/Au (5/45 nm) layer is overlaid onto the Al$_2$O$_3${ }/SiO$_2$/Ti/Au layer and the TMD crystal to create the contacts of the devices (the overlay region is clearly visible in Fig. 1e). The same two steps enable the fabrication of $5\:\mathrm{\mu m}\times5\:\mathrm{\mu m}$ Pt reference electrodes placed within $150\:\mathrm{\mu m}$ of the TMD crystal, specifically designed to measure the potential close to the device (as shown in Fig. 1e; in this case SiO$_2$ is used to provide electrical insulation and to determine the location where the reference electrodes contact the LICGC substrate). Except for the FETs whose transistor characteristics are shown in Fig. 2e-f and Fig. S7, for which the LASPT glass-ceramic was used, the devices were fabricated using the AG-01 glass-ceramic.  \\

%For clarity we state that we do not place a SiO$_2$ insulating layer below the reference electrodes used in FETs where the SiO$_2$ passivation layer is absent below the Pt contacts; it is not required because in such devices V$_{Ref}$ saturates for $V_\mathrm{G}>0.75\:\mathrm{V}$ independently of the size and location of the reference electrode.

\textbf{Ionic-liquid gated FETs}
Details regarding the fabrication of the ionic-liquid gated FETs can be found in the main text and in our earlier works\cite{Spectroscopy_2012,LET_2014_M,artificial_2018,MoS2_2018,CrossQuantum_2019,ILS_review_2021}. We note however that the devices studied here do not use a PMMA window to confine the ionic liquid to the channel of the device. As a consequence there is a significant capacitive coupling between the Pt contacts and the ionic liquid, resulting in the absence in Fig. 2b. of the $I_\mathrm{G}$ peaks caused by the onset of hole and electron accumulation.  \\

\textbf{Cyclic voltammetry devices}
The CV-GC devices consist of $700\:\mathrm{\mu m}\times700\:\mathrm{\mu m}$ metal electrodes (either Pt/Au (10/40 nm), SiO$_2$/Pt/Au (40/10/40 nm) ) directly deposited on the LICGC via ebeam-lithography and ebeam-evaporation. For the CV-IL devices, the same lithography procedure was used to deposit $700\:\mathrm{\mu m}\times700\:\mathrm{\mu m}$ 50 nm thick Au electrodes. The ionic liquid (DEME-TFSI) is drop casted as a final step.\\

\textbf{Electronic transport measurements}
The room-temperature measurements are performed under high vacuum ($1\times 10^{-6}\:\mathrm{mbar}$) in a continuous flow-cryostat (Oxford instruments), and the low-temperature measurements under dilute Helium environment in a Teslatron Cryogenfree II cryostat. The DC FET and CV measurements are performed via  Keithley 2400 source/measure units and digital multimeters (Keysight 34401A and Agilent 34410A) coupled to home-made voltage/current source/measure amplifiers. The AC measurements are performed in current or voltage bias mode using a SRS830 lock-in amplifier (Stanford Research Systems) coupled to the same home-made amplifiers. The gate voltage applied to the LICGC gate can be removed below 150 K without any observable effect on the conductance of the FETs. The devices can therefore be safely disconnected to change the measurement configuration (e.g., DC to AC or voltage to current bias).\\

\medskip
\textbf{Supporting Information} \par %Please delete the Suppporting Information statement if it is not applicable. Please supply Supporting Information in another file. Supporting information should not be provided in .tex format
Supporting Information is available from the ancillary file or from the author.\\

% Acknowledgements
\medskip
\textbf{Acknowledgements} \par %delete if not applicable))
The authors gratefully acknowledge Alexandre Ferreira for continuous and precious technical support. A.F.M. gratefully acknowledges financial support from the Swiss National Science Foundation (Division II) and from the EU Graphene Flagship project.

% References
\medskip

\bibliography{refs.bib}

\end{document}